\documentclass[prb,twocolumn,groupedaddress,superscriptaddress,showpacs,floatfix]{revtex4-1}
\usepackage{amsfonts}

\usepackage[dvips]{graphicx}
\usepackage{dcolumn}
\usepackage{bm}
\usepackage{amsmath}
\usepackage{amssymb}
\usepackage{revsymb4-1}
\usepackage{hhline}

\begin{document}

\title{Light propagation in tunable exciton-polariton one-dimensional photonic crystals}
\author{E. S. Sedov}
\email[Electronic address: ]{evgeny\_sedov@mail.ru}
\affiliation{School of Physics and Astronomy, University of Southampton, SO17 1NJ Southampton, United Kingdom}
\affiliation{Department of Physics and Applied Mathematics, Vladimir State University named after A. G. and N. G. Stoletovs, Gorky str. 87, 600000, Vladimir, Russia}
\author{E. D. Cherotchenko}
\affiliation{School of Physics and Astronomy, University of Southampton, SO17 1NJ Southampton, United Kingdom}
\author{S. M. Arakelian}
\affiliation{Department of Physics and Applied Mathematics, Vladimir State
University named after A. G. and N. G. Stoletovs, Gorky street 87, 600000, Vladimir, Russia}
\author{A. V. Kavokin}
\email[Electronic address: ]{a.kavokin@soton.ac.uk}
\affiliation{School of Physics and Astronomy, University of Southampton,
SO17 1NJ Southampton, United Kingdom} \affiliation{CNR-SPIN, Viale del Politecnico 1, I-00133, Rome, Italy}
\affiliation{Spin Optics Laboratory, St. Petersburg State University, Ul'anovskaya 1, Peterhof, St. Petersburg 198504, Russia}

\begin{abstract}
Simulations of propagation of light beams in specially designed multilayer semiconductor structures (one-dimensional photonic crystals) with embedded quantum wells reveal characteristic optical properties of resonant hyperbolic metamaterials.
A strong dependence of the refraction angle and the optical beam spread on the exciton radiative lifetime is revealed.
We demonstrate the strong negative refraction of light and the control of the group velocity of light by an external bias through its effect upon the exciton radiative properties.
\end{abstract}

\maketitle

\section{Introduction}

Metamaterials are artificial composite structures that demonstrate unusual optical properties  not achievable in natural materials~\cite{OptMetamatShalaevBook2010,NatPhoton79582013}.
Hyperbolic metamaterials (HMMs) represent an important group of metamaterials characterised by specific relations between components of dielectric permittivity ($\varepsilon$) and magnetic susceptibility ($\mu$) tensors~\cite{NatPhoton79582013}.
Namely, the diagonal components of either $\varepsilon $ or $\mu $ tensors in
HMMs have opposite~signs.

The presently studied HMMs are mostly based on metal-dielectric composites,
whereas the presence of specially designed metallic elements, \mbox{\emph{e.~g.}}
metallic spheres, disks, rods, etc., provides the required properties
of $\varepsilon $ and $\mu $ tensors~\cite{LaserPhysLett,NatPhoton1412007,PhysRevLett.90.077405,apl8422442004}.
The successful application of this approach to the fabrication of HMMs has been
demonstrated in the microwave spectral range with use of metal wires and
split ring resonators~\cite{PhysRevLett.76.4773,Shelby77}.
The metal-dielectric HMMs have been developed in the near-infrared and visible frequency bands, see \mbox{\emph{e.~g.}}~\cite{Science31516862007,ApplPhysLett99151115,SchonbrunIEEEPhotonTechnolLett171196,JOptics140630012012}.
One of the intriguing effects observed in HMMs is negative refraction~\cite{OptExpress16154392008,Shalaev:05,NatPhoton1412007,Shelby77}.
This effect is very promising for the development of hyperlenses~\cite{Science31516862007,OptExpress1482472006} characterized by the spatial resolution beyond the diffraction limit as well as in optical cloaking~\cite{Science314977,NatPhoton12242007}.

As noticed in~\cite{OptExpress16154392008}, negative refraction materials can be based not only on HMMs but on photonic crystal (PC) structures as well.
Although both HMMs and PCs are complex structured materials, unlike the former, which can be considered as a quasi-homogeneous medium with the effective constitutive parameters, the latter possesses the elementary building blocks being of the same size order as the impinging wavelength.
Various optical effects in periodically stratified media that are, in fact, one-dimensional PC's caused by anomalous refraction  have been observed in works of Russell (see \textit{e.g.} Refs.~\onlinecite{ApplPhysB392311986,PRA333232,RussellBirksINPhotonicBandGapMaterials1996}).
It has been shown in Ref.~\onlinecite{PhysRevB.62.10696} that the hyperbolic dispersion of light modes can be achieved in such structures with positive $\varepsilon $ and $\mu$ tensors.
Recent publications ~\cite{Kavokin2005387,PhysRevLett.93.073902,NatureMater69462007} provide further details on the design of pure dielectric PC-based materials with hyperbolic dispersion.

Both in the metal-dielectric metamaterials and in the PC-based materials the propagation of
light is governed by the structure parameters and composition, so that the
external control of light trajectories is hardly achievable.
Meanwhile, it is technologically important to provide a significant tunability of the optical properties of the  materials, \emph{e.~g.} using ultrashort optical pulse pumping of free carriers~\cite{PhysRevB.66.081102}, introducing small perturbations in the near field of the photon modes or temperature tuning of the refractive index~\cite{ApplPhysLett848462004}.
To answer
this technological challenge one may try to embed in PCs objects that
willingly respond to the external impact and help modifying the optical
properties of the whole system~\cite{Tomadin:10}. The role of such tunable
elements can be played  \mbox{\emph{e.~g.}} by ultracold two-level atoms~\cite%
{PhysRevA.89.033828,Jaksch200552}, quantum dots~\cite{Nature4458962007},
diamond nitrogen-vacancy centers~\cite{Su:08}, or Cooper-pair boxes.
%Further improvement of tunability of HMMs requires the use of tunable elements that are robust at high temperatures.

Recently, specially designed planar multilayer semiconductor structures demonstrating properties of resonant HMMs (RHMMs) have been theoretically proposed.
The model structure considered in~\cite{Sedov2015} represents a modified semiconductor Bragg mirror, containing periodically arranged quantum wells (QWs).
It was shown by modeling that such a structure should demonstrate properties of HMMs in a spectral range where the dispersion of its optical eigen-modes acquires a hyperbolic character.
Based on this similarity of eigen-mode dispersion properties, we refer to such a structure as RHMM.
This structure may be also qualified as a PC because its period is on the order of a half wavelength of light.
We shall use the term HMM because it was introduced in the previous publication~\cite{Sedov2015} and it correctly describes the phenomenology  of light propagation in the considered structures.
The structure proposed in Ref.~\onlinecite{Sedov2015} has a significant advantage over traditional HMMs since it contains no metallic elements that would absorb and
scatter light leading to unavoidable losses. In addition, it offers the
possibility of tuning its optical properties by applying external electric
and magnetic fields that modify the exciton radiative life-time and,
consequently, affect the exciton-light coupling strength.

In this paper, we demonstrate how the proposed RHMM allows tailoring of the wave packets and control of the effective refractive index and the group velocity of light.
We model propagation of light beams in such RHMMs and show the negative refraction and the control of the trajectory of the beams by external bias.

%==================================================
%==================================================
%==================================================
%==================================================

\begin{figure}[tbp]
\centering
\includegraphics[width=0.6\columnwidth]{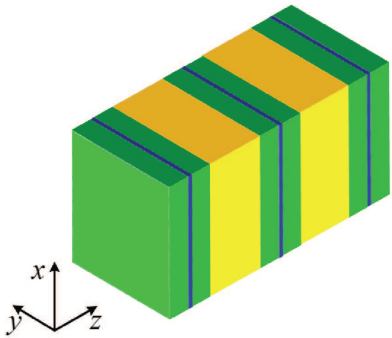}
\caption{(Color online) Schematic picture of RHMM, a spatially-periodic array of alternating dielectric layers of different widths and refractive indices, with the layers of one type contain single QWs placed in their centers.}
\label{FIG_Structure}
\end{figure}

\section{Negative effective mass in RHMM: the Low Angle of Incidence Limit}

We consider the structure
schematically shown in Fig.~\ref{FIG_Structure} which was first proposed by some
of the authors of this paper in~\cite{Sedov2015}. The structure represents
a spatially-periodic array of alternating dielectric layers of different
widths and refractive indices, with the layers of one type containing
single QWs placed in their centers.
We assume cylindric symmetry of the system and introduce the in-QW-plane radial coordinate~$\rho$.

To describe the optical dispersion properties of the structure, we use the
transfer matrix technique following \cite{AnnPhys3501948,kavokin2007}. Namely, we introduce
a vector $\Phi = \left(E ({\mathbf{r}, t}), c B ({\mathbf{r}, t}) \right)^{%
\mathrm{T}} $, where $E ({\mathbf{r}, t})$ and $B ({\mathbf{r}, t})$ are the
amplitudes of electric and magnetic fields, a superscript ($^{\mathrm{T}}$) denotes transposition.
Considering propagation in the $z$ direction that coincides with the structure growth axis, it is possible to link the field amplitude of a
light wave entering the layer of width $D$ with one of a light wave leaving this layer by the following equation:
\begin{equation}
\hat{T} \left. \Phi \right| _{z=z_i} = \left. \Phi \right| _{z=z_i + D},
\end{equation}
where $\hat{T}$ is the transfer matrix through the period of the structure.
Since each period $D$ is formed by four successive layers, namely, the first
type half layer, QW, the first type half layer again, and the second type layer,
the transfer matrix $\hat{T}$ is found as a product of the transfer matrices
through the corresponding layers in the reverse order, \emph{i.e.}, $\hat{T}
= \hat{T}_{d_2} \hat{T}_{d_1 / 2} \hat{T}_{\mathrm{QW}} \hat{T}_{d_1/2}$.
The submatrices are given by
\begin{subequations}
\label{TrMatrices}
\begin{eqnarray}
\label{TrMatricesa}
& &\hat{T}_{d_j} =
\begin{pmatrix}
\cos (k_{zj} d_j) & \frac{i k_0}{k_{zj}} \sin (k_{zj} d_j) \\
\frac{i k_{zj}}{k_0} \sin (k_{zj} d_j) & \cos (k_{zj} d_j)%
\end{pmatrix},
\\ \label{TrMatricesb}
& &\hat{T} _{\mathrm{QW}} =
\begin{pmatrix}
1 & 0 \\
2\frac{k_{z1} r_{\text{QW}} }{k_0 t_{\text{QW}}} & 1%
\end{pmatrix}%
,
\end{eqnarray}
\end{subequations}
where $k_{0}=\omega /c$, and $k_{z1,2}=\sqrt{n_{1,2}^{2}k_{0}^{2} - {k}%
_{\rho }^{2}}$, and $\mathbf{k}_{\rho }=(k_{x},k_{y})$ is the in-plane
wavevector component; $n_{1,2}$ are the refractive indices of the first and
second sublayers respectively.
In the calculations we assume QWs to be infinitely thin.
The coefficients $r_{\text{QW}}$ and $t_{\text{QW}} = 1 + r_{\text{QW}}$ are the amplitude reflection and transmission coefficients of the~QW.
According to \cite{kavokin2007,IvchenkoBook}, the former is given by
\begin{equation}
\label{ReflectionWithSuscept}
r_{\text{QW}} = \frac{\left. i n_1 k_0 \Gamma _0 \right/ k_{z1}}{\omega_{X} - \omega -i \left( \Gamma + \left. n_1 k_0 \Gamma _0 \right/ k_{z1}\right) }.
\end{equation}
where $ \Gamma _{0}$ and $ \Gamma$ determine the radiative and nonradiative decay rates, respectively, and $\omega _{X}$ is the QW exciton resonance frequency.
The relative QW permittivity in the growth direction can be estimated with respect to the following expression:
\begin{equation}
\label{PermittGrDir}
\varepsilon _{\text{QW}} =
\varepsilon_{1} \left(1 + \frac{2 n_1 \Gamma _{0} / k_{z1} D}{\omega_{X} - \omega -i \Gamma} \right).
\end{equation}

\begin{figure}[tbp]
\centering
\includegraphics[width=\columnwidth]{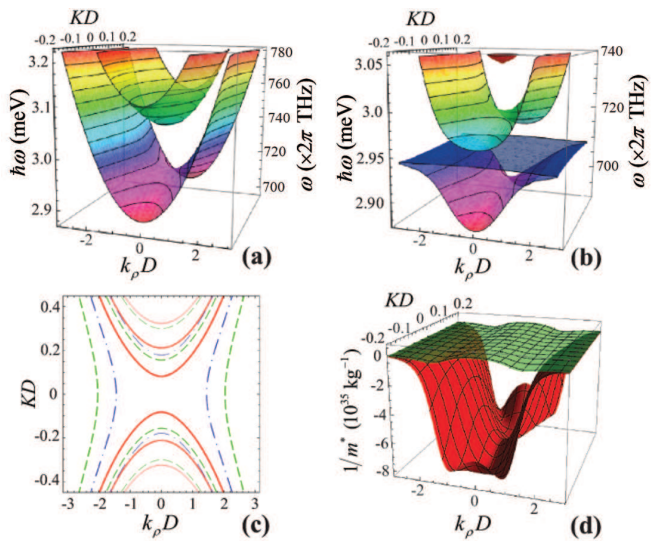}
\caption{(Color online) Dispersion of the photonic eigen-modes for the structure (a) without and (b) with embedded QWs (b).
(c) --- Equi-frequency contours in the reciprocal space showing the structure eigen modes  belonging to the lowest dispersion band corresponding to the different values of $\Gamma_0$:
$\Gamma_0 = 0$ for the red (solid) curves,
$\hbar \Gamma_0 = 2$~meV for the green (dashed) curves,
and $\hbar \Gamma_0 = 10$~meV for the blue (dash-dotted) curves.
Different thicknesses of the curves correspond to different energies $\hbar \omega$ (from the thickest to the thinnest): 2.94~meV, 2.89~meV and 2.84~meV, respectively.
(Table I contains the corresponding relative QW permittivities $\varepsilon _{\text{QW}}$.)
(d) Inverse exciton-polariton effective mass tensor components in the structure.
The red surface corresponds to the effective mass in the  $z$-direction,  $m ^{*} _{z}$ ($z$~being the growth axis of the structure), and the green surface corresponds to the in-plane effective mass, $m ^{*} _{\protect\rho}$.
The parameters used in the calculation are given in the text.
The QW radiative decay rate for (d) is taken as $\hbar \Gamma _0 = 2\,\text{\rm meV}$. }
\label{FIG_Dispersions}
\end{figure}

The dispersion equation for the eigen-modes of an infinite periodic structure is given by
\begin{equation}  \label{InitialDispEq}
\cos (KD) = \frac{1}{2} \text{Tr}[\hat{T}],
\end{equation}
where $K$ is a normal to QW-plane pseudo-wave vector component. Figures~\ref{FIG_Dispersions} (a) and (b) demonstrate dispersions of the light modes in a modified Bragg mirror
structure without [Fig.~\ref{FIG_Dispersions}(a)] and with [Fig.~\ref{FIG_Dispersions}(b)] embedded periodically arranged narrow QWs
characterized by the radiative decay rate $\hbar \Gamma _{0} = 2\,\text{\rm meV}$.
As a model structure we consider a $\text{GaN}/ \text{Al}_{0.3}\text{Ga}_{0.7}\text{N}$
Bragg mirror with embedded thin $\text{In}_{0.12}\text{Ga}_{0.88}\text{N}$ quantum wells.
The thicknesses of the layers and their refractive indices are taken as $%
d_{1}=64.8\,\text{\rm nm}$, $n_{1}=2.55$ and $d_{2}=115.3\,\text{\rm nm}$, $n_{2}=2.15$, the period of
the lattice $D=d_{1}+d_{2}$ is $180.1\,\text{\rm nm}$.
For the given parameters the structure exhibits a second photonic band gap centered to $\hbar \omega _{B} \simeq 3\,\text{\rm eV}$ in a QW-free case, see.~Fig.~\ref{FIG_Dispersions}(a).
The QW exciton resonance energy $\hbar \omega _{X}$ is tuned close to the lower boundary of the second photonic band gap, $\hbar \omega _{X} \simeq 2.95\,\text{\rm eV}$.
The QW nonradiative decay rate is taken as $\hbar \Gamma = 0.1\,\text{\rm meV}$.
The radiative decay rate $\Gamma _0$ is a tunable parameter that strongly depends on the applied electric field.
The tuning of $\Gamma _0$ by the external bias is addressed in the Appendix.

The presence of QWs leads to two principal changes in the dispersion of the eigen-modes.
The first one is the appearance of four dispersion branches instead of the two characteristic of a QW-free structure due to the
vacuum field Rabi-splitting originated from the QW exciton-photon coupling.
The resulting eigen-states of the system are exciton-polaritons.

The other change is the formation of the three-dimensional polaritonic band gap [see Fig.~\ref{FIG_Dispersions}(b)].
It is necessary to mention, that the presence of QWs pushes the lowest dispersion branch (LB) to the lower energies and the greater the shift, the larger the value of $\Gamma _0$.
If a QW exciton is resonant with an eigenmode of the photonic cavity structure, the exciton-photon results in the appearance of new exciton-polariton eigen modes, in the strong coupling regime.
In the limit of weak coupling, exciton and photon modes would cross each other.
In contrast, the strong coupling manifests itself in the anticrossing (avoided crossing) of the modes.
The dispersion of the structure eigenmodes is no more photonic or excitonic, but polaritonic.
The increase of the exciton-photon coupling due to the increase of the exciton radiative decay rate results in the growing energy level repulsion.
The exciton radiative decay rate is controlled by the applied electric field (see the Appendix).

Figure~\ref{FIG_Dispersions}(c) demonstrates equi-frequency contours (EFCs) in the $( K k_{\rho} )$ plane for a number of LB eigen energies $\hbar \omega$  for different values of $\Gamma_{0}$.
It is clearly seen, that with the increase of $\Gamma_0$ opposite branches of EFCs approach each other until the gap in the $K$ direction closes and the gap in the $k_{\rho}$ direction opens.
Table I below gives values of the relative QW permittivity $\varepsilon _{\text{QW}}$ in the growth direction estimated with respect to Eq.~\eqref{PermittGrDir} for $\Gamma_0$ and $\omega$ used in Fig.~\ref{FIG_Dispersions}(c).
The case $\Gamma_0 = 0$ corresponds to the absence of QWs in the structure, hence the permittivity of $\text{GaN}$ layers remains unmodulated and equal to $\varepsilon _1$.

\begin{table}
\caption{Relative QW permittivity $\varepsilon_{\text{QW}}$ in the growth direction.}
\begin{center}
\begin{tabular}{|c|c||c|}
\hline
& $\hbar \Gamma_0 = 2$~meV &  $\hbar \Gamma_0 = 10$~meV  \\
\hline\hline
$\hbar \omega $ & $\varepsilon _{\text{QW}} $ & $\varepsilon _{\text{QW}}$  \\ \hline
2.84~eV & $6.59372 + i 0.00008$ & $6.95861 + i 0.00041$\\
2.89~eV & $6.66685 + i 0.00027$ & $7.32424 + i 0.00137$\\
2.94~eV & $7.47172 + i 0.00969$  & $11.34860 +i 0.04846$\\
\hline
\end{tabular}
\end{center}
\end{table}

Hereafter, we consider only the LB and neglect three upper polariton branches that are split in energy and do not affect the propagation of light in the spectral range of our interest.
Our structure behaves like a HMM in the specific frequency range in the vicinity of the LB saddle point $(K, k_{\rho} = 0)$.

\subsection{A QW-free modified Bragg mirror}

Let us first consider the QW-free structure assuming $\Gamma_{0} = 0$.
Following~Ref.~\onlinecite{Kavokin2005387}, we can easily obtain the saddle-point frequency $\omega _0$.
To do this, we restrict ourselves to the limit $K,k_{\rho} \ll 1 / D$, \emph{i.e.}, we consider the system in the vicinity of the center of the first Brillouin zone (BZ).
We make the Taylor expansion of Eq.~\eqref{InitialDispEq} and obtain in the zeroth order
\begin{equation}  \label{DispRelEqnNearKkEqZero}
1=\cos \left( \theta _1 \right) \cos \left( \theta _2 \right) - \frac{n_1 ^2
+ n_2 ^2}{2n_1 n_2} \sin \left( \theta _1 \right) \sin \left( \theta _2
\right),
\end{equation}
where $\theta _{1,2} = \omega _{0j} n_{1,2} d_{1,2} / c$.
The subscript $j$ numerates dispersion branches.
For the QW-free structure $j = 1,2$.
In the general form  the effective photonic mass tensor components are given by $m_{j,\alpha}^{*}= \hbar \left( \left. \partial ^2 \omega _j \right/ \partial k_{\alpha}
^2\right) ^{-1}$ with $k_{\alpha} = K, k _{\rho}$.
Taking the second derivative of both right and left parts of Eq.~\eqref{InitialDispEq} over the wave vector components in the saddle point it is easy to obtain  analytical expressions for the effective photonic mass tensor components:
\begin{widetext}

\begin{subequations}
\label{PhotMasses}
\begin{multline}
\label{transPhotMass}
\left. m_{ j, z} ^{\rm ph}  \right| _{K,k_{\rho} = 0}= \frac{\hbar }{D^2 c} \left[
\sin(\theta _2) \cos(\theta _1) \left( d_2 n_2 + \frac{d_1 (n_1^2 + n_2^2)}{2n_2} \right)+
\sin(\theta _1) \cos(\theta _2) \left( d_1 n_1 + \frac{d_2 (n_1^2 + n_2^2)}{2n_1} \right)
\right],
\end{multline}
\begin{multline}
\label{inPlanePhotMass}
\left. m_{ j, \rho} ^{\rm ph} \right| _{K,k_{\rho} = 0} = \frac{D^2 \omega _{0j}}{c} m_{\rm j, z} ^{\rm ph}
\left[
\sin (\theta _2)\cos (\theta _1) \left( \frac{d_2}{n_2} + \frac{d_1 (n_1^2 + n_2^2)}{2 n_1 ^2 n_2} \right)  \right. \\
 \left. +
\sin (\theta _1)\cos (\theta _2) \left( \frac{d_1}{n_1} + \frac{d_2 (n_1^2 + n_2^2)}{2 n_1  n_2^2} \right) -
\sin{\theta _1}\sin{\theta _2} \frac{c(n_1^2 - n_2^2)^2}{2 n_1^3 n_2^3 \omega _{0j}}
\right] ^{-1},
\end{multline}
\end{subequations}
\end{widetext}
where we take $\theta _{1,2} = \theta _{1,2}|_{\omega_j = \omega_{0j}}$.

At $k _{\rho} \ll 1/D$ a photonic band gap appears at the $K$ direction.
Let us find a half-width of the band gap in the structure without embedded QWs following the method described in the Supplemental Material to Ref.~\onlinecite{Sedov2015}.
We introduce the parameter $\zeta = \frac{n_1 d_1}{n_2 d_2} - 1$ that characterizes the relative optical path lengths in the structure layers.
Generally speaking, for a modified Bragg structure this parameter is close to zero.
The center of the photonic band gap is characterized by a Bragg frequency that according to~\cite{PhysRevA.64.013809} is given by $\omega _{B} = \left. 2 \pi c \right/ (n_1 d_1 + n_2 d_2)$.
We introduce the parameter $\delta = \omega_{01} / \omega _{B} - 1$ that is also small in comparison with 1.
We expand Eq.~\eqref{DispRelEqnNearKkEqZero}  up to the second order in $\zeta$ and $\delta$.
As a result, we obtain the following expression for $\delta$: $\delta = \pm \frac{(n_1 - n_2) \zeta}{2(n_1 + n_2)}$.
Now it is easy to find a half-width of the band gap as~$\Omega _{B} = \omega _{B} |\delta|$.

In accordance with the foregoing, the eigenfrequency of the lower photonic branch at $K,k _{\rho} = 0$ is given by
\begin{equation}
\label{LowBrEigFrInCent}
 \omega _{01} \simeq \omega _B - \Omega _B.
\end{equation}

\subsection{Modified Bragg mirror with embedded QWs}

When adding the QWs to the structure, a new term describing the exciton impact appears in the right-hand side of the Eq.~\eqref{DispRelEqnNearKkEqZero}:
\begin{multline}
\frac{i r_{\text{QW}} }{t_{\text{QW}}} \left\{ \cos \left(
\theta _2 \right) \sin \left( \theta _1 \right) \vphantom{\frac{n_1}{n_2}} \right.\\
\left. + \left[
\frac{n_1}{n_2} \cos ^2 \left( \frac{\theta _1}{2}\right) \right. \left. -
\frac{n_2}{n_1} \sin ^2 \left( \frac{\theta _1}{2} \right) \right] \sin
\left( \theta _2 \right) \right\}.
\end{multline}

The expressions for the effective exciton-polariton masses in the vicinity of the saddle point can be obtained by the same technique as described above in the form
\begin{subequations}
\label{PolEffMasses}
\begin{eqnarray}
\left. m ^{*} _{j,z} \right| _{K,k_{\rho} = 0} &=& m ^{\mathrm{ph}} _{j,z} +i \Theta _1, \\
\left. m ^{*} _{j,\rho} \right| _{K,k_{\rho} = 0} &=& \frac{D^2 \omega }{c} m_{j,z} ^{*} \left(\Theta _2 - i \Theta _3 \right)^{-1}.
\end{eqnarray}
\end{subequations}

The parameters $\Theta _{1,2,3} (\omega _0, \Gamma _0)$ that characterize the QW exciton impact on the optical
properties of the considered RHMM are found in the form
\begin{widetext}
\begin{subequations}
\begin{multline}
\Theta _1 = \frac{\hbar r_{\text{QW}} ^{(0)}}{c D^2 t_{\text{QW}} ^{(0)}}
\left\{
i \frac{c r_{\text{QW}} ^{(0)}}{\Gamma _0 t_{\text{QW}} ^{(0)}}
\left[
\cos (\theta _2) \sin (\theta _1) + \sin (\theta _2) \left( \frac{n_1}{n_2} \cos ^2 \left( \frac{\theta _1}{2} \right) - \frac{n_2}{n_1} \sin ^2 \left( \frac{\theta _1}{2} \right) \right)
\right] \right. \\
 \left.+
\left[
d _1 n_1 \cos (\theta _1) \cos (\theta _2) - \sin (\theta _1) \sin (\theta _2) \left( d_2 n_2 +\frac{d_1 (n_1 ^2 + n_2 ^2)}{2 n_2} \right) \right. \right.
 \left. \left. +d_2 n_2 \left( \frac{n_1}{n_2} \cos ^2 \left( \frac{\theta _1}{2} \right) -\frac{n_2}{n_1} \sin ^2 \left( \frac{\theta _1}{2} \right) \right) \cos (\theta _2)
\right]
\right\},
\end{multline}
\begin{multline}
\Theta _3 = \frac{c^2 r_{\text{QW}} ^{(0)}}{t_{\text{QW}} ^{(0)}} \left[
\sin (\theta _1) \sin (\theta _2) \left( \frac{d_2}{n_2} + \frac{d_1 (n_1 ^2 + n_2 ^2)}{2 n_2 n_1 ^2} \right) -
\frac{d_1}{n_1} \cos (\theta _1) \cos (\theta _2) + \frac{c}{n_1 ^2 \omega _{0j}} \cos(\theta _2) \sin (\theta _1) \right.\\
\left. + \frac{c}{n_1 n_2 \omega _{0j}} \left( \frac{n_1 ^2}{ n_2 ^2} \cos ^2 \left( \frac{\theta _1}{2} \right) + \sin ^2 \left( \frac{\theta _1}{2} \right) \left( 1-\frac{2 n_2 ^2}{n_1 ^2} \right) \right) \sin (\theta _2) - \frac{d_2}{n_1} \left( \frac{n_1 ^2}{n_2 ^2} \cos ^2 \left( \frac{\theta _1}{2} \right) - \sin ^2 \left( \frac{\theta _1}{2} \right) \right) \cos(\theta _2)
\right],
\end{multline}
\end{subequations}
\end{widetext}
the parameter $\Theta _2$ is the rectangular bracket in the right part of~Eq.~\eqref{inPlanePhotMass};
$r_{\text{QW}} ^{(0)} \equiv \left. r_{\text{QW}} \right| _{K,k_{\rho} = 0}$,
$t_{\text{QW}} ^{(0)} \equiv \left. t_{\text{QW}} \right| _{K,k_{\rho} = 0}$.

In the vicinity of the saddle point of LB, the effective mass tensor
components $m_{\rho, \,z} ^{*} \equiv m_{1, \rho, \,z} ^{*}$ have opposite signs. This is clearly seen
in~Fig.~\ref{FIG_Dispersions}(d) where the dependencies of the inverse
in-plane (green surface), $m^{*} _{\rho}$, and transverse (red surface), $%
m^{*} _{z}$, effective masses on the position in the 1st BZ are shown. One
can see that $m^{*} _{\rho}>0$ while $m^{*} _{z}<0$. It also should be
mentioned that for the considered model structure the absolute value of $%
m^{*} _{z}$ is at least one order of magnitude smaller than $m^{*} _{\rho}$.
For example, according to Eqs.~\eqref{PolEffMasses}, the ratio $| \left.  m _{\rho } ^{*}\right/ m _{z} ^{*}|$ at the saddle point is  about 20.1 for QW-free structure, 21.6 for the structure with embedded QWs with $\hbar \Gamma _{0} = 2$~meV and grows to 30.7 for the structure with $\hbar\Gamma _{0} = 10$~meV.
Such a big difference introduces a strong anisotropy to the optical properties of the considered structure.

%==================================================
%==================================================
%==================================================
%==================================================

\begin{figure}[t!]
\centering
\includegraphics[width=0.85\columnwidth]{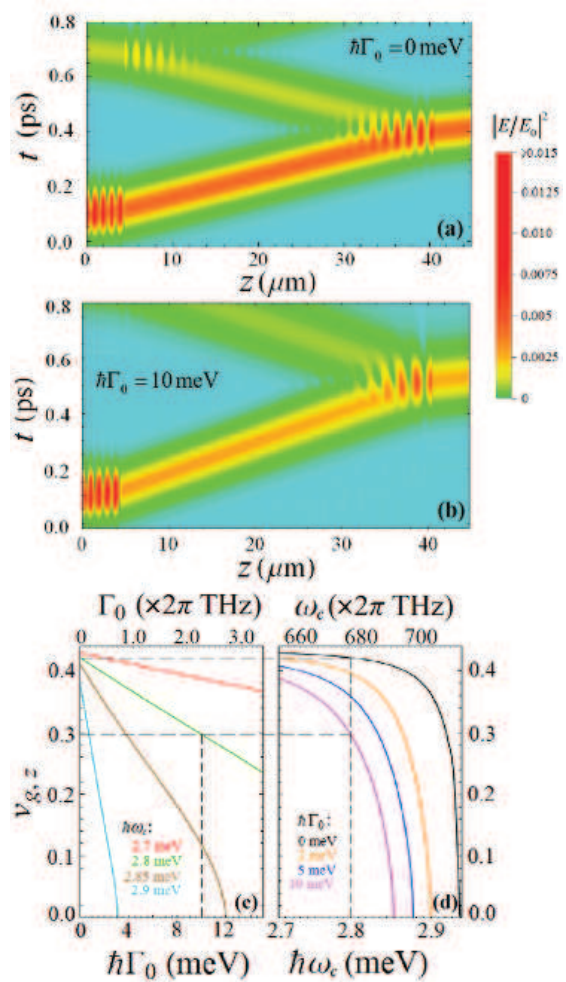}
\caption{(Color online) Femtosecond laser pulse propagation in the multilayer structure schematically shown in Fig.~\ref{FIG_Structure}.
The parameter $\hbar \Gamma _{0}$ is taken as (a) $0\,\text{\rm meV}$ (that is equivalent to
the absence of QWs in the structure) and  (b) $10\,\text{\rm meV}$.
Graphs (c) and (d) demonstrate the parametric dependencies of the group velocity of light
in the $z$ direction, $v _{g, \, z}$, (c) on $\Gamma _{0}$ for a number of fixed values of  the wave-packet central frequency component $\omega _c$  and (d) on~$\omega_c$ for different values of $\Gamma _0$ (d) with $k _{\protect\rho} = 0$.
Values of $v _{g, \, z}$ are given in units of the speed of light in vacuum~$c$.
The vertical dashed lines correspond to  $\Gamma _0$ in (c) and $\omega _c$ in (d) from (a) and (b).
Horizontal dashed lines indicate the group velocities of the
wave packet,  with the considered values of $\Gamma _0$. }
\label{FIG_Pulse_Propagations}
\end{figure}

\section{Light speed manipulation in RHMM}

First, let us discuss the
propagation of a femtosecond laser pulse in the growth direction of the
structure ($z$-axis). We consider the Gaussian pulse in the form
\begin{equation}
E(z,t)=E_{0}\exp \left[ -\frac{(t-t_{0})^{2}}{2t_{w}^{2}}\right] \exp
[-i\omega _{c}t]\exp [-ik_{z}z],
\end{equation}%
centered on the frequency $\omega _{c}$; $E_{0}$ determines the pulse amplitude, $t_{w}$ is the half-width duration of the
pulse. Here we consider the wave packet whose spatial width $\rho _{w}$
exceeds the in-plane structure size, and we assume the intensity of light to
be uniformly distributed in the QW plane in each layer. We consider the normal
incidence geometry.
In the numerical calculations we take $t_{w}=50\,\text{\rm fs}$ and
$t_{0}=0.1\,\text{\rm ps}$, $\hbar \omega _{c}=0.95 \times \hbar \omega _{X}\simeq 2.8\,\text{\rm eV}$; $k_{\rho }=0$.

Figures~\ref{FIG_Pulse_Propagations} (a) and (b) demonstrate light pulse
propagation in the structure calculated for different values of $\Gamma _{0}$.
In Fig.~\ref{FIG_Pulse_Propagations}(a) we take $\Gamma _{0}=0$, which corresponds to the QW-free Bragg mirror.
In Fig.~\ref{FIG_Pulse_Propagations}(b) we consider the case of a Bragg mirror with embedded QWs characterized by a high radiative decay rate, $\hbar \Gamma _{0}=10\,\text{\rm meV}$.
Propagation of light has been modelled in the system starting with a vacuum layer of width $25\times D$ on the left-hand side of the structure.
In the middle part of the system we have placed an RHMM of $200$ layers of width $%
D$ each. The right-most part represents a $25\times D$ thick vacuum layer
again. It is clearly seen that $v_{g,\,z}$ significantly decreases with the
increase of $\Gamma _{0}$.
It is also confirmed by~Figs.~\ref{FIG_Pulse_Propagations} (c) and~\ref{FIG_Pulse_Propagations}(d) demonstrating the dependence of $v_{g,\,z}$ on $\Gamma _{0}$ (c)  for the fixed values of $\omega _c$ and its
dependence on $\omega _c$ (d) for several fixed values of $\Gamma
_{0}$. Such a tendency can be qualitatively explained as follows.
Once the parameter $\Gamma _{0}$ increases, the lowest branch moves down in energy, see~Figs.~\ref{FIG_Dispersions} (a) and~\ref{FIG_Dispersions}(b). Since in the
vicinity of $K,k_{\rho }\simeq 0$ the dependence $\omega (K)$ for the lowest
branch is convex, to conserve the energy the wave packet should reduce its
wavevector and group velocity $v_{g,\,z}$, see EFCs in Fig.~\ref{FIG_Dispersions}(c).
It is important to mention that
this conclusion is only correct in a specific frequency range, namely, for $\omega_c < \omega _{0}$.

The regular optical patterns in~Figs.~\ref{FIG_Pulse_Propagations} (a) and~\ref{FIG_Pulse_Propagations}(b) describe the interference of the propagating pulse and the pulses
reflected from vacuum-RHMM and RHMM-vacuum interfaces.

%==================================================
%==================================================
%==================================================
%==================================================

\section{Negative Refractive Response of RHMM}

\begin{figure}[tbp]
\centering
\includegraphics[width=0.85\columnwidth]{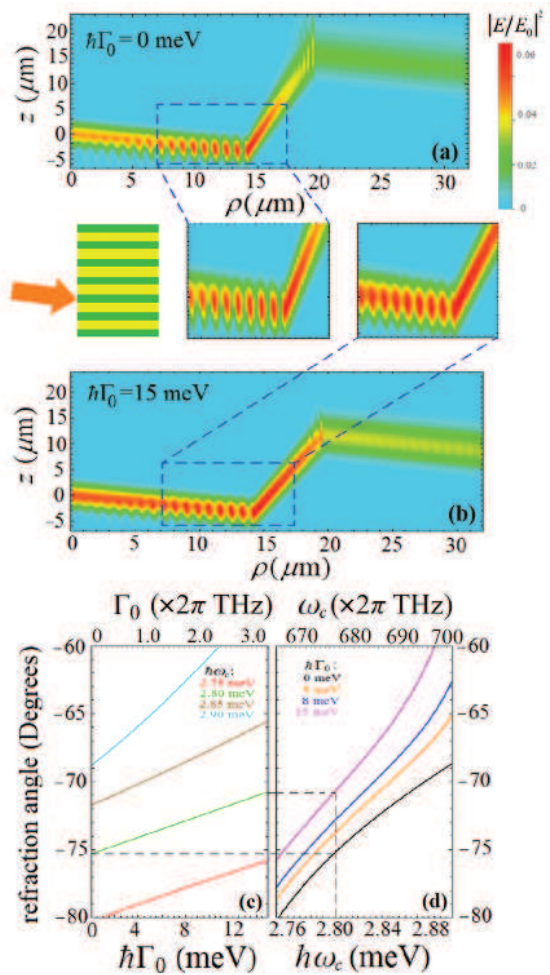}
\caption{(Color online)  Focalized light beam propagation in the $\protect\rho z$ plane simulated for
different values of $\Gamma _0$. The parameter $\hbar \Gamma _{0}$ is taken as $0\,\text{\rm eV}$ for (a) and $15\,\text{\rm meV}$ for (b). Panels (c) and (d) show the dependencies of the
refraction angle on $\Gamma _{0}$ for a number of fixed values of $\omega_c$ and on $\protect\omega_c$ for different values of $\Gamma _0$ with $K _{0} = 0.6 D^{-1}$.
The vertical dashed line on (d) corresponds to the value of $ \omega _c$ taken from  (a) and (b).
Horizontal lines indicate the refraction angles of the beams for the considered values of $\Gamma _0$. }
\label{FIG_Distribution_MANY}
\end{figure}

Let us now consider a different
geometry of the experiment, where a monochromatic spatially focalized light
beam enters the structure from its side and propagates in the $\rho z$ plane;
see the inset in Fig.~\ref{FIG_Distribution_MANY}. We consider the
transmission of light through the interface between vacuum and RHMM. The
medium on the left is a vacuum characterized by a familiar linear dispersion $%
\omega =ck_{\mathrm{vac}}$. The spatially modulated structure of a Bragg
mirror we now consider in the continuous approximation as a homogeneous
effective dielectric medium characterized by the dispersion given by Eq.~%
\eqref{InitialDispEq} and by the tensorial effective mass of photons. This
approximation is valid if the typical spatial size of the light beam (beam
width $z_{w}$) is much larger than period of the structure, $z_{w}\gg D$, and at sufficiently low incidence angles.

Now we consider the propagation of a monochromatic Gaussian light beam of
frequency $\omega _{c}$
\begin{equation}
E(z, \rho, t)=E_0 \exp \left[ -\frac{(z-z_{0})^{2}}{%
2z_{w}^{2}}\right] e^ {-i(K_{0}z + k_{\rho} \rho)} e^{-i\omega_{c}t},  \label{GaussBeamEq}
\end{equation}%
where $z_{w}$ is the beam spatial width. The beam propagates at an oblique angle to
the structure, that is set by $K_{0}$. In our simulations we take $z=0$, $%
z_{w}=12\times D$, and $\omega _{c}=0.95\omega _{X}$. The wave vector component $%
K_{0}$ is taken as $K_{0}D=0.6$. To model the propagation of light in the $\rho z
$ plane of the structure, we use the transfer matrix technique adapted for
the new geometry.
We check the accuracy of this numerical procedure in
limiting cases by analytical calculations realised in the effective photonic
mass approximation using the expressions~\eqref{PolEffMasses} for the effective mass tensor components in the vicinity of the saddle point.

Figure~\ref{FIG_Distribution_MANY} demonstrates the light beam
propagation in the $\rho z$ plane of the structure without [Fig.~\ref{FIG_Distribution_MANY}(a)] and with [Fig.~\ref{FIG_Distribution_MANY}(b)]
embedded QWs in the regime of a high radiative decay rate $\hbar \Gamma _{0}=15\,\text{\rm meV}$.
We considered the model RHMM of $30\times D$ width limited by vacuum on both sides.

One can see from Figs.~\ref{FIG_Distribution_MANY} (a) and~\ref{FIG_Distribution_MANY}(b) that the light beam undergoes the negative refraction  in the considered geometry.
Moreover, since the absolute value of the effective mass in the $z$-direction is one order of magnitude larger than the in-plane mass, the negative refraction appears to be very strong.
Clearly, the advantage of polaritonic RHMM over dielectric PC structures is in the suitability for the external
control of the refraction angle in a range of several degrees that is achieved just by tuning
the value of $\Gamma _{0}$.
This tuning can be done, \mbox{\emph{e.~g.}} by application of the external bias (see the Appendix).
Figures~\ref{FIG_Distribution_MANY} (c) and~\ref{FIG_Distribution_MANY}(d) demonstrate the dependencies of the refraction angle on $\Gamma _{0}$ (c) for the fixed values of $\omega_c $ and on $\omega_c $ (d) for the fixed values of~$\Gamma _{0}$.
One can see that the absolute value of the
refraction angle decreases with the increase of $\Gamma _{0}$. This tendency
is maintained as long as the frequency of the beam $\omega _{c}$ is less
than $\omega _{0}$. If this condition is violated, some values of $k_{\rho }$
in the vicinity of the first BZ center become forbidden, which affects the
shape and angular dispersion of the beam.

We also note that the parameter $\Gamma _{0}$ significantly affects the
spread of the optical beam. Comparing Figs.~\ref{FIG_Distribution_MANY} (a)
and~\ref{FIG_Distribution_MANY}(b), it can be concluded that the increase of $\Gamma _{0}$ up to a
certain limit reduces the light beam blurring. This is also correct only in the
limit of $\omega _{c}<\omega _{0}$. Finally, we want to mention that Figs.~%
\ref{FIG_Distribution_MANY} (a) and~\ref{FIG_Distribution_MANY}(b) also illustrate qualitatively the
influence of $\Gamma _{0}$ on the transmission properties of RHMMs. The
variation of transmittivity affects interference patterns in the vicinity of
the surfaces of the considered structure.

\section{Conclusions}
We have considered a planar RHMM based on a modified Bragg mirror with embedded periodically arranged QWs.
The optical properties of this RHMM are tunable by changing the radiative decay rate of embedded quantum wells.
The latter can be done by application of external electric and magnetic fields due to their strong influence on the exciton oscillator strength (see the Appendix).
This enables one to control the group velocity and propagation direction of light as well as its spatial distribution.

This work was supported by the Russian Foundation for Basic
Research Grants No. 16-32-60104, No. 15-59-30406, and No. 15-52-52001, by grant
of President of Russian Federation for state support of young Russian
scientists No. MK-8031.2016.2, by the Russian Ministry of Education and
Science state tasks No. 2014/13, 16.440.2014/K and by the EPSRC Hybrid
Polaritonics Programme grant.

\appendix*
\section{Impact of the external electric field on the exciton radiative decay rate}

Here we discuss the mechanism of tuning of $\Gamma _{0}$ by the external bias.
It is known that the radiative exciton lifetime $\tau _{\text{rad}}$ in a quantum well $z$ direction is governed  by the overlap of the electron and hole wave functions $\psi _{e} (z)$ and  $\psi _{h} (z)$ and the exciton in-plane Bohr radius~$a_{B}$ (Refs.~\onlinecite{kavokin2007,IvchenkoBook}).
In general,  $\tau _{\text{rad}}$ is larger, the smaller  the overlap integral $\left[ \int \psi _{e} (z) \psi _{h} (z) dz \right]^2$.
The following expression links these parameters~\cite{SolovevKukushkin2006,IvchenkoBook}:
\begin{equation}
\label{Gamma0onOverlapIntegral}
\Gamma _{0} = \frac{1}{2 \tau_{\text{rad}}} = \overline{\Gamma} _{0} \left[ \int \psi _{e} (z) \psi _{h} (z) dz \right]^2,
\end{equation}
where $\overline{\Gamma} _{0} = \left. {\sqrt{\varepsilon_B} \omega _{LT}  k_0 a_{B} ^3} \right/ (a_{B} ^{\text{2D}}) ^2$ is the quantity of the appropriate dimension,
$\varepsilon _{B}$ is a background dielectric constant,
$\omega _{LT}$ is the longitudinal-transverce splitting frequency,
$a_{B}$ and $a_{B} ^{\text{2D}}$ are the exciton Bohr radii in the bulk and in QW, respectively.
The electron-hole overlap integral is strongly sensitive to the applied electric field normal to the QW plane due to the quantum confined Stark effect, see \emph{e.~g.} Ref.~\onlinecite{PhysRevB.63.035315}.
Generally speaking, both $a_{B} ^{\text{2D}}$ and the overlap integral in~\eqref{Gamma0onOverlapIntegral} depend on the applied electric field, however this dependence for the former is negligibly weaker than for the latter and can be ignored.

\begin{figure}[t!]
\centering
\includegraphics[width=0.85\columnwidth]{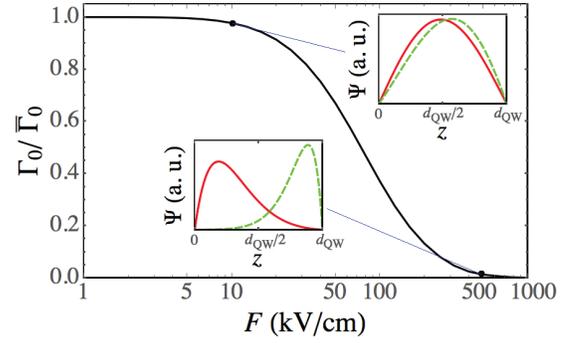}
\caption{(Color online) Squared electron-hole overlap integral in dependence on the applied external electric field.
The latter is given in logarithm scale.
The inserts demonstrate schematically the wave functions of an electron (red solid curves) and a hole (green dashed curves) in QW at specified values of $F$ that are 10~kV/cm (for the lower insert) and 500~kV/cm (for the upper insert).
}
\label{FIG_OverlapInt}
\end{figure}

To estimate the overlap integral, one should solve Schr\"{o}dinger equations for both  electron $ \psi _e (z)$ and hole $ \psi _h (z)$ envelope functions associated  with the eigenenergies $E_{e,h}$:
\begin{equation}
\label{ElHolSchEq}
\left[-\frac{\hbar ^2}{2 m _{i} ^{*}} \frac{\partial ^2}{\partial z^2} + V_i (z) + q_i F z  \right] \psi_i (z) = E_i \psi _i (z),
\end{equation}
where $i = e,h$;
$m_{i} ^{*}$ is the effective mass;
$V_{i} (z)$ is the potential for the $i$-th carrier that in the simplest case we take equal to zero inside QW and to infinity outside.
$F$ is the stationary external field applied in the growth direction, and
$q_i$ determines the charge of the $i$-th carrier.

Next, we use the variational approach to retrieve electron and hole wave functions.
We take trial functions in the form~\cite{PikusFTP1992}
\begin{subequations}
\begin{eqnarray}
\psi _{e} (z) &=& A_{e} \sin \left( \frac{ \pi z }{ d_{\text{QW}}} \right) e ^{ - \alpha _{e} \frac{|z|}{d_{\text{QW}}} },\\
\psi _{h} (z) &=& A_{h} \sin \left( \frac{ \pi (d_{\text{QW}} - z) }{ d_{\text{QW}}} \right) e ^{  - \alpha _{h} \frac{|d_{\text{QW}} - z|}{d_{\text{QW}}}}, \qquad
\end{eqnarray}
\end{subequations}
where $A_{e,h} = A_{e,h} (\alpha_{e,h})$ are normalization parameters determined from the orthonormality condition $\int_{-\infty} ^{\infty} |\psi _{e,h}|^2 d z = 1$.
$\alpha _{e,h}$ are the only variational parameters that are found by minimization of the carrier energy
\begin{equation}
E_i = \langle \psi _i| H_i | \psi _i \rangle = \int_{-\infty} ^{\infty} \psi _i ^{*} (z) H_i  \psi _i (z) d z ,
\end{equation}
where $H_i$ represents the Hamiltonian corresponding to Eq.~\eqref{ElHolSchEq}.
For the estimations we take the carriers' effective masses $m_{e}  ^{*}= 0.2 m_0$ and  $m_h ^{*}= 0.8 m_0$ with $m_0$ being the mass of a free electron, and the QW width $d_{\text{QW}} = 10$~nm.
Figure~\ref{FIG_OverlapInt} shows the squared electron-hole overlap integral as a function of the applied electric field.
It is clearly seen that for the strong fields exceeding 1000~kV/cm the overlap integral is sufficiently small.
At the same time, for the fields less than 1~kV/cm it remains unchanged.
These estimates define the tunability range of the applied field that allows one to manipulate~$\Gamma _0 $.

It is necessary to mention that the theoretical curve presented in Fig.~\ref{FIG_OverlapInt} does not reflect the true dependence of the estimated parameters on the external field.
A number of additional effects that have not been taken into account in the simulations can modify the dependence quite considerably.
For example, the external field applied in the QW plane direction leads to spatial separation in electrons and holes and it also leads to change of the overlap integral, see~Ref.~\onlinecite{PRB3210431985}.
Another possible effect is associated with the screening of the electric field by the counteracting  field generated by free carriers.
The partial cancellation of the internal field impact with the increase the free carrier's density have been discussed in~\cite{ApplPhysLett8347912003,ApplPhysLett1012311072012}.
In contrast with thin (on the order of 1~nm width) QWs, in the thicker (10~nm) GaN-based QWs, the carriers are spatially separated due to internal electric fields.
In the case of small free carrier's density a large electric field indeed induces a large spatial separation between electrons and holes, leading to a long recombination lifetime.
The change of the radiative lifetime as a result of the interplay between a built-in electric field inside the quantum well and a small external electric field  is discussed in~\cite{ApplPhysLett942111072009}.
On the contrary, when the density of carriers increases, this leads to the enhancement of the induced electric field.
The screening effects of the electric field due to carriers become important.
They lead to the increase of the overlap integral and the decrease of the recombination lifetime as a result.
In this case, the maximum on the dependence in Fig.~\ref{FIG_OverlapInt} appears for large enough values of $F$.
To take into account the screening effect, an additional term $ q_{i} \Phi _{i} (z) z \psi _{i} (z)$ should be included in Eq.~\eqref{ElHolSchEq}, where $\Phi_i (z)$ depends on the carrier densities $|\psi _i|^2$, see~Refs.~\onlinecite{PikusFTP1992, PhysRevB.63.035315%,PhysStatusSolidiB2529402015
}.

Another important factor affecting the  dependence of $\Gamma _0$ on the external field is temperature.
The temperature impact on the recombination lifetime was considered in, \emph{e.~g.}~\cite{ApplPhysLett6919361996,PhysStatusSolidiB2162911999}.
The authors concluded from the photoluminescence intensity measurements that the radiative lifetime linearly increases with temperature, which also modifies the specified dependence.

Last but not least, although we did our calculations for GaN-based QWs of 10~nm width, the initial choice of the QW width allows us to pick up the reference value of $\Gamma _{0}$ in a wide range as well.
To illustrate the remarkable dependence of the radiative lifetime on a QW width we refer to~\cite{PhysStatusSolidiB2162911999,PhysRevB.64.121304} where this problem is the focus of attention.

%The expression linking the photoluminescence lifetime $\tau _{\text{PL}}$ and the radiative recombination lifetime $\tau_{\text{rad}}$ is following:
%$\tau _{\text{PL}} ^{-1} = \tau _{\text{rad}} ^{-1} + \tau _{\text{nonrad}} ^{-1} + \tau _{\text{trans}}^{-1}$, where $\tau _{\text{nonrad}}$ and $\tau _{\text{trans}}$ represent the nonradiative lifetime and the lifetime of carrier transfer from the luminescence area.
%The latter term is negligibly small as a rule.
%In this regard, we can estimate the radiative recombination lifetimes based on the observations of  the external impact dependent change in the PL lifetime as $\tau _{\text{rad}} = \left. \tau _{\text{PL}} \right/ \eta _{\text{int}}$ where $\eta _{\text{int}} = \left. \tau _{\text{nonrad}} \right/ (\tau _{\text{rad}} + \tau _{\text{nonrad}})$ is the internal quantum efficiency.
%The problem of PL lifetime change in presence of applied electric field is in focus of the works~\cite{ApplPhysLett942111072009,ApplPhysLett8347912003,PhysStatusSolidiB2529402015}.

%More accurate estimates of the radiative decay rate in dependence of the applied electric field are a subject of our further research

\bibliography{Microcavities}

%\begin{thebibliography}{99}
%\bibitem{RMP7115911999} G. Khitrova and H. M. Gibbs, F. Jahnke, M. Kira, and S. W. Koch, Rev. Mod. Phys. \textbf{71}, 1591 (1999).

%\bibitem{PhysRevB640453162001} N. H. Kwong, R. Takayama, I. Rumyantsev, M. Kuwata-Gonokami, and R. Binder, Phys. Rev. B \textbf{64}, 045316 (2001).

%\bibitem{OptCommun2853732012} F. Ungana, U. Yesilgul, E. Kasapoglu, H. Sari, I. S\"{o}kmen,  Opt. Commun. \textbf{285}, 373 (2012).

%\bibitem{Kavokin} A.V. Kavokin, J.J. Baumberg, G. Malpuech, F.P. Laussy, "Microcavities", Oxford University Press, Oxford, 2007.

%\end{thebibliography}

\end{document}